\documentclass[aps,pra,showpacs,eqsecnum,twocolumn]{revtex4}
\usepackage{graphicx}
\usepackage{amsmath,amssymb,amsthm,dsfont,bm}
\usepackage{color}
\usepackage{soul}
\usepackage{ulem}

\begin{document}
\preprint{}
\title{Classical and quantum complementarity}
\author{Raquel Galazo}
\author{Irene Bartolom\'e} 
\author{Laura Ares}
\author{Alfredo Luis}
\email{alluis@fis.ucm.es}
\homepage{http://www.ucm.es/info/gioq}
\affiliation{Departamento de \'{O}ptica, Facultad de Ciencias
F\'{\i}sicas, Universidad Complutense, 28040 Madrid, Spain}
\date{\today}

\begin{abstract}
Two complementary observables can be measured simultaneously so that the exact individual distributions can be recovered by a proper data inversion. We apply this program to the paradigmatic example of the Young interferometer from the classical and quantum points of view. We show complete parallelism between complementarity in the quantum and classical theories.  In both domains, complementarity manifests in a pathological behavior for the inferred joint distribution. 
\end{abstract}


\maketitle

\section{Introduction}

Complementarity is a fundamental idea originally introduced in the quantum domain. In the standard approach is formulated as the impossibility of a joint simultaneous determination of two observables.  Although  complementarity  is usually understood to be a pure quantum phenomenon,  this is not quite so, and complementarity holds also in the classical domain \cite{L8,QVE17,QVE18,EQV17}. This is another example of phenomena that were originally believed to be of quantum origin, but that can be found also in classical optics, as it is the case of the Zeno effect, for example \cite{GZ,LS691,LS692,SL,YIK,PLG,GPL,MKKW19}. 

In this work we demonstrate a complete parallelism between complementarity in quantum and classical optics. For definiteness, we focus on  the most seminal example of path-interference complementarity: the Young interferometer. Complementarity will manifest in the attempt to derive a joint distribution for such two complementary variables.

Our starting point is that two complementary observables can be always observed simultaneously, also in the quantum domain, provided that the observation is imprecise enough \cite{L1,L2,L3,L4,M1,M2,M3,M4,S1,GY88,S3,JSV95,BE96,S6,S7,BK98,BST99,OS99,DR00,L6,L7,CKW14}. In our case, the joint observation will be allowed by marking the light at each aperture by a different polarization state. Then, the interference is observed while keeping track of the polarization state that contains the path information.

But even if the observation is imprecise, it may provide complete and exact information about the two variables in question, that can be then extracted via a suitable data inversion procedure. This inversion procedure will be applied to the imprecise, simultaneous observation of the amount of light at the apertures and the interference pattern, addressing the existence of a noiseless joint distribution for these observables. The idea is that complementarity will manifest in some kind of pathology for such attempted joint distribution. 

The main result we find is that this will occur in classical optics in exactly the same terms as in quantum optics. The only quantum-classical difference is the physical meaning of the distribution reflecting complementarity: simple light intensity in the classical domain versus photon probability in the quantum case. Otherwise, a distinctive feature of the analysis presented in this work is the use of exactly the same tools in both domains.

We focus on the classical-optics domain in Sec. 2, showing the negativity in the inferred joint distribution. We address the quantum domain in Sec. 3, showing that it closely follows the classical complementarity. The basis of the inversion procedure is presented in Appendix A.

\bigskip
 
\section{Classical domain}

\subsection{Settings. Exact, unobserved scenario}

Our physical system is a standard Young interferometer with two small apertures which are illuminated by a monochromatic wave. The apertures will be labelled by the index $z = \pm 1$. The field at the apertures will be denoted as $E_z$, with light intensities 
\begin{equation}
\label{IR}
I_Z (z) = \langle | E_z |^2 \rangle ,
\end{equation}
where the angular brackets denote ensemble averages. 

Interference is observed in the far field leading to the intensity 
distribution $I_\Phi (\phi)$ in the appropriate units, 
\begin{equation}
\label{Ip}
 I_{\Phi}(\phi) = \frac{1}{2 \pi} \langle\vert E_{1} + E_{-1}e^{-i\phi}\vert^2\rangle  ,
 \end{equation}
this is to say 
\begin{eqnarray}
\label{IP}
& I_{\Phi}(\phi) =  \frac{1}{2 \pi} \left [ I_Z (1)  + I_Z (-1)  \right . & \nonumber \\
& \left . + 2 | \mu | \sqrt{I_Z (1)I_Z (-1)}\cos(\phi + \delta) \right ] , &
\end{eqnarray}
where $\phi$ is the phase difference acquired from the slits to the observation point, $\mu$ is the complex degree of coherence and $\delta$ its phase
\begin{equation}
\label{mu}
\mu = | \mu | e^{i \delta} =  \frac{\langle E_1E_{-1}^*\rangle }{\sqrt{\langle\vert E_1\vert^2\rangle\langle\vert E_{-1} \vert^2\rangle}} .
\end{equation}

 Without loss of generality, we normalize intensities so that 
\begin{equation}
\label{norm}
 I_Z (1)+ I_Z (-1) = \int_{2 \pi} d \phi I_\Phi (\phi) =1. 
\end{equation}

In a more formal way, the complete field information is presented by the $2 \times 2$ cross-spectral density matrix, with matrix elements $\Gamma_{i,j}= \langle E_i E_j^\ast \rangle$:
\begin{equation}
\label{Ga}
\Gamma =  \begin{pmatrix} \langle | E_1 |^2 \rangle & \langle  E_1 E_{-1}^\ast \rangle \\ \langle  E_1^\ast E_{-1} \rangle & \langle | E_{-1} |^2 \rangle \end{pmatrix} ,
\end{equation}
this is 
\begin{equation}
\Gamma = \begin{pmatrix} I_Z (1)  & \mu \sqrt{I_Z (1)I_Z (-1)} \\ \mu^\ast \sqrt{I_Z (1)I_Z (-1)} & I_Z (-1)  \end{pmatrix} .
\end{equation}

\subsection{Hilbert-space description of classical interference}

Before proceeding, let us formalize the classical scenario in the same formal language of quantum mechanics to gain insight and parallel the quantum scenario. The field state at the apertures is a complex two-dimensional space, 
 
\begin{equation}
\begin{pmatrix}  E_1 \\ E_{-1} \end{pmatrix} = E_1 \begin{pmatrix}  1 \\ 0 \end{pmatrix} + E_{-1}  \begin{pmatrix}  0 \\ 1 \end{pmatrix} 
\end{equation}
which can be expressed in the traditional quantum ket notation

\begin{equation} 
\label{vE}
\begin{pmatrix}  E_1 \\ E_{-1} \end{pmatrix} \equiv | E \rangle = E_1 | 1 \rangle + E_{-1} | -1 \rangle, 
\end{equation}
being
\begin{equation}
\label{|r>}
|  z = 1 \rangle = \begin{pmatrix}  1 \\ 0 \end{pmatrix}, \qquad | z=  -1 \rangle = \begin{pmatrix}  0 \\ 1 \end{pmatrix} .
\end{equation}
This is that we can describe light state at the apertures and their combination with the tools of a two-dimensional Hilbert space $\mathcal{H}_s$.

In this regard, the very same cross-spectral density matrix $\Gamma$ has the properties of a quantum density matrix $\rho$ : it is Hermitian, positive semidefinite and after the normalization of intensities we have also $\mathrm{tr} \Gamma = 1$. This is the mixed-state counterpart of the pure state expression  (\ref{vE}) when the complex amplitudes $E_{\pm 1}$ present the characteristic randomness of partial coherence. More especifically  
\begin{equation}
\label{ra}
\Gamma = \bigg  \langle | E \rangle \langle E | \bigg \rangle ,
\end{equation}
where care must be taken to distinguish the quantum-like bra-ket notation with the angles expressing the typical classical-optics averages in Eqs.  (\ref{IR}),  (\ref{Ip}),  (\ref{mu}),  (\ref{Ga}), and (\ref{ra}). 
\color{black}

With all this, our basic complementary quantities $I_Z (z), I_\Phi (\phi) $ can be expressed in a pure quantum-mechanical fashion as
\begin{equation}
\label{prob}
I_Z (z) = \langle z | \Gamma | z \rangle, \qquad I_\Phi (\phi)  = \langle \phi | \Gamma | \phi \rangle,
\end{equation}
where  the states representing interference $| \phi \rangle$ are the {\it phase states} \cite{B1,B2,P1,P2,P3,P4,P5,P6}
\begin{equation}
\label{ps}
| \phi \rangle = \frac{1}{\sqrt{2 \pi}} \begin{pmatrix} 1 \\ e^{i \phi} \end{pmatrix} .
\end{equation}

\subsection{Joint observation and inversion}

The goal of applying the inversion procedure is to obtain a joint distribution $I (z,\phi)$, meaning the amount of light leaving the aperture $z$ in the 
direction specified by $\phi$, so that the true intensities (\ref{IR}) and (\ref{IP}) are its marginals 
\begin{equation}
I_Z (z) = \int_{2\pi} d\phi I (z,\phi) , \quad I_{\Phi}(\phi) = \sum_{z=1,-1} I(z, \phi).
\end{equation}
To perform a simultaneous observation we transfer information of one of the observables, say $I_Z (z)$, to additional degrees of freedom, for example the polarization state, so that $I_Z (z)$ will be inferred from polarization measurements, while $I_\Phi (\phi)$ will be determined from the intensity distribution at the observation screen disregarding polarization. The observation of the interference keeping track of the polarization provides a joint distribution $\tilde{I}(p,\phi)$, where $p$ refers to the orientation of a polarizer, to which we will apply the inversion procedure explained in Appendix A to obtain $I(z,\phi)$.

Polarization can be suitably described by a two-dimensional Hilbert space $\mathcal{H}_p$ with two orthogonal basis vectors $| \rightarrow \rangle$ and $| \uparrow \rangle$, representing horizontal and vertical linear polarization for example. The light illuminating the slits has horizontal polarization  $| \rightarrow  \rangle$. Then, the transfer of information about $I_Z (z)$ to the polarization state can be easily achieved in practice by placing  a half-wave plate in one of the apertures performing the transformation of the polarization state from $| \rightarrow  \rangle$ to $\cos \theta | \rightarrow \rangle+ \sin \theta | \uparrow \rangle$ for some angle $\theta$.  

Now the scenario has been enlarged including the polarization degrees of freedom, so the complete setting is described by the Hilbert space $\mathcal{H}_s \otimes \mathcal{H}_p$. Then, the field state on the apertures including polarization is now described by the $4 \times 4$ cross-spectral density matrix 
\begin{equation}
\label{tG}
\tilde{\Gamma }= \left \langle | \tilde{E} \rangle \langle  \tilde{E} | \right \rangle, \quad | \tilde{E} \rangle = E_1 | 1 \rangle | u_1 \rangle + E_{-1} | -1 \rangle | u_{-1} \rangle ,\end{equation}
where 
\begin{equation}
  |u_1\rangle = \cos \theta | \rightarrow \rangle+ \sin \theta | \uparrow \rangle, \qquad |u_{-1}\rangle =  | \rightarrow  \rangle  .
\end{equation}
On the plane where we observe the interference now we place an ideal polarizator. We record the intensity  $\tilde{I} (p, \phi)$ for two orthogonal orientations of the polarizator axis denoted with the index $p=\pm 1$ represented by the unit vectors in the polarization space $\mathcal{H}_p$ 
\begin{eqnarray}
\label{|p>}
& | p=1 \rangle = \cos \vartheta | \rightarrow \rangle + \sin \vartheta | \uparrow \rangle , & \nonumber \\
 & & \\
 & | p = - 1 \rangle = - \sin \vartheta | \rightarrow \rangle + \cos \vartheta | \uparrow \rangle , & \nonumber 
\end{eqnarray}
where $\vartheta$ is an arbitrary angle. As shown in appendix A of Ref. \cite{LU13}, the optimum choice holds for $2 \vartheta - \theta = \pi/2$. To gain insight, and without loss of generality, we can consider such a case. Thus, the recorded intensity in the interference plane is
\begin{equation}
\label{jo}
\tilde{I} \left ( p, \phi \right ) =  \langle p | \langle \phi | \tilde{\Gamma } | \phi \rangle | p \rangle ,
\end{equation}
leading to 
\begin{eqnarray}
& \tilde{I}\left(1,\phi\right) = \frac{1}{2\pi} \bigg [ \sin^2 \vartheta  I_Z(1) +  \cos^2\vartheta  I_Z(-1) & \nonumber \\
& + |\mu |\sin \left( 2 \vartheta \right) \sqrt{I_Z(1)I_Z(-1)} \cos\left(\phi+\delta\right) \bigg ]   , & \nonumber \\ & & \\
& \tilde{I}\left(-1,\phi\right) =  \frac{1}{2\pi} \bigg [ \cos^2 \vartheta  I_Z(1) +  \sin^2\vartheta  I_Z(-1) & \nonumber \\
& + |\mu |\sin\left(2 \vartheta \right) \sqrt{I_Z(1)I_Z(-1)} \cos\left(\phi+\delta\right) \bigg ]  . & \nonumber
\end{eqnarray}

The corresponding marginals for polarization and interferometric intensity are 
\begin{equation}
\tilde{I}_{P} \left( p \right) = \int_{2\pi} d\phi \tilde{I}(p,\phi) , \quad \tilde{I}_{\Phi}(\phi)  = \sum_{p=\pm 1} \tilde{I}(p, \phi) ,
\end{equation}
with 
\begin{eqnarray}
\label{mP}
& \tilde{I}_{P}\left(1\right)  =  \sin^2 \vartheta I_Z(1) +  \cos^2\vartheta I_Z(-1) , & \nonumber \\ & &  \\
& \tilde{I}_{P}\left(-1\right) =  \cos^2 \vartheta I_Z(1) + \sin^2\vartheta I_Z(-1), & \nonumber
\end{eqnarray}
or in matrix form 
\begin{equation}
\label{matrix}
\begin{pmatrix} \tilde{I}_P (1) \\ \tilde{I}_P (-1) \end{pmatrix} =
\begin{pmatrix} \sin^2 \vartheta  & \cos^2\vartheta  \\ \cos^2 \vartheta & \sin^2\vartheta \end{pmatrix}
\begin{pmatrix} I_Z (1) \\  I_Z (-1) \end{pmatrix} ,
\end{equation}
while for the phase marginal distribution we have
\begin{eqnarray}
& \tilde{I}_{\Phi}(\phi) =  \frac{1}{2\pi} \bigg [ I_Z(1) + I_Z(-1)  & \nonumber \\
& + 2 | \mu |\sin \left ( 2 \vartheta \right ) \sqrt{I_Z(1)I_Z(-1)}\cos\left(\phi+\delta\right) \bigg ] & .
\end{eqnarray}

The idea is that the polarization measurement $\tilde{I}_{P} \left( p \right) $ represents the imperfect observation of $I_Z (z)$, while 
$\tilde{I}_{\Phi}(\phi) $ is an imperfect observation of $I_{\Phi}(\phi) $. Therefore, we have to look for the inverting functions $M_{Z,\Phi}$ that carry out the inversions 
\begin{eqnarray}
& I_Z (z) = \sum_{p=\pm 1} M_Z (z,p) \tilde{I}_P(p) , & \nonumber \\
 & & \\
 &I_{\Phi}\left(\phi\right) = \int_{2\pi} d \phi^\prime M_{\Phi}(\phi,\phi^\prime )\tilde{I}_{\Phi}( \phi^\prime ) . & \nonumber
\end{eqnarray}
These are, in matrix form for the slit-polarization pair, as the inverse of (\ref{matrix}), 
\begin{equation}
\begin{pmatrix} I_Z(1) \\ I_Z(-1) \end{pmatrix} =
\begin{pmatrix} M_Z(1,1) & M_Z(1,-1) \\ M_Z(-1,1) & M_Z(-1,-1) \end{pmatrix}
\begin{pmatrix} \tilde{I}_P(1) \\  \tilde{I}_P(-1) \end{pmatrix} ,
\end{equation}
with 
\begin{equation}
\label{Mz}
M_Z (z,p)=\frac{1}{\cos \left ( 2 \vartheta \right )}
\begin{pmatrix}
-\sin^2\vartheta  & \cos^2\vartheta  \\
\cos^2 \vartheta & -\sin^2\vartheta 
\end{pmatrix} ,
\end{equation}
while for the interferometric-phase variable 
\begin{equation}
\label{Mf}
M_{\Phi}(\phi,\phi^\prime ) = \frac{1}{2\pi} \left( 1 + \frac{2}{\sin \left ( 2 \vartheta \right )}\cos\left(\phi - \phi^\prime \right)\right) .
\end{equation}

Then we extend the inversion procedure to the joint distribution as
\begin{equation}
I(z,\phi) = \sum_{p = \pm 1}\int_{2\pi} d  \phi^\prime   M_Z (z,p) M_{\Phi}(\phi,  \phi^\prime  )\tilde{I}(p,  \phi^\prime ) ,
\end{equation}
to obtain
\begin{equation}
\label{fr}
I (z, \phi) =  \frac{1}{2 \pi} \left [ I_Z (z) + | \mu | \sqrt{I_Z (1)I_Z(-1)} \cos (\phi + \delta) \right ].
\end{equation}

\subsection{Pathology}

\begin{figure}[h]
    \centering
    \includegraphics[width=8cm]{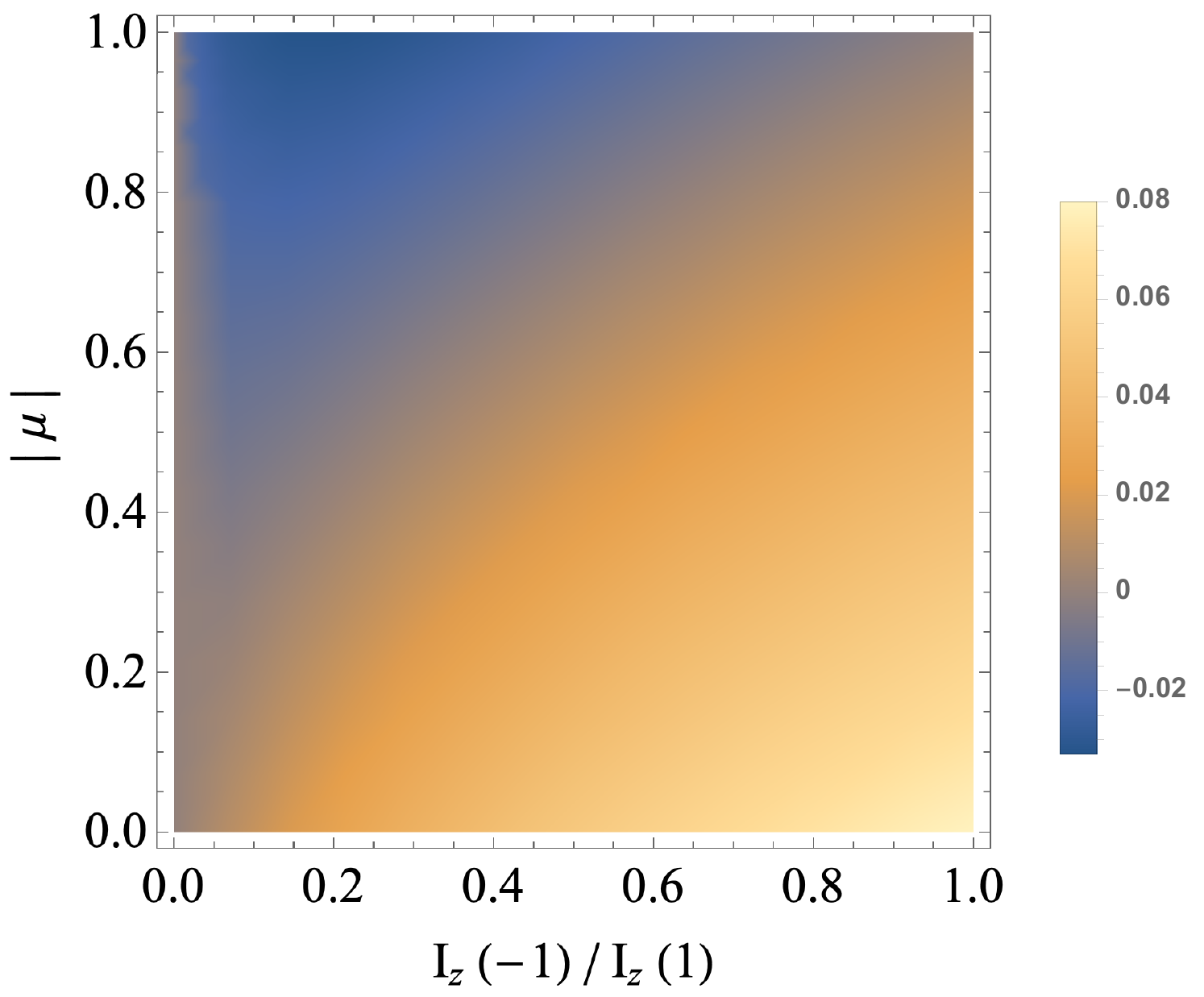}
    \caption{Retrieved joint intensity distribution showing negative values under condition (\ref{PathologyClass}), assuming $I_Z (-1) < I_Z(1)$ without loss of generality.}
    \label{figIPath}
\end{figure}

The pathological results we are looking for hold when the inverted distribution (\ref{fr}) takes negative values, $I(z,\phi)<0$, for some particular values of $z$ and $\phi$. To this end, let us find the minimum value in Eq. (\ref{fr}) by choosing simply $\phi= - \delta + \pi$, so that  $\cos \left(\phi + \delta \right) = -1$. Then we have either $I(1,\phi) <0$ or $I(-1,\phi) <0$ provided that 
\begin{equation}
\label{PathologyClass}
| \mu |^2  > \min{\{ I_Z(1) /I_Z (-1), I_Z(-1) /I_Z (1)\} }. 
\end{equation}{}
This is clearly illustrated in Fig. \ref{figIPath}. 
{}

It is worth noting that the pathological behaviour requires some threshold value for the degree of coherence.

\section{Quantum domain}

\subsection{Settings. Exact, unobserved scenario}

Focusing in the case of just one photon, the density matrix is 
\begin{equation}
\rho  =  \frac{1}{2} \left ( \sigma_0 +  \bm{s} \cdot \bm{\sigma} \right ) ,
 \end{equation}
where $\sigma_0$ is the $2 \times 2$ identity, $\bm{\sigma}$  are the Pauli matrices, and $\bm{s} = \mathrm{tr} (\rho \bm{\sigma} )$ is a three-dimensional real vector with $| \bm{s} | \leq 1$.  Within this context the observable 
$Z$ is represented by the third Pauli  matrix $\sigma_z$, with eigenstates $ | z = \pm 1 \rangle$, with $z=1$ meaning that the photon is found in one of the apertures and $z=-1$ in the other aperture. The corresponding probabilities  are
\begin{equation}
\label{PZe}
P_Z (z) =  \langle z | \rho | z \rangle= \frac{1}{2} \left (1 + z s_z \right ) ,
\end{equation}
clear counterpart of the classical $I_Z(z)$.

As shown for example in the classical equation (\ref{prob}), in this simplified context interference equals phase distribution. 
The quantum description of phase-like variables is a rather tricky point since there is no simple operator for the phase or phase difference \cite{B1,B2,P1,P2,P3,P4,P5,P6}. Maybe the best quantum description of relative phase suited for our purposes is given by a positive-operator-valued measure. More specifically, the phase statistics $P_\Phi (\phi)$ is given by projection of the photon state on the nonorthogonal phase states  in Eq. (\ref{ps}) \cite{P3}, this is
\begin{equation}
\label{ps2}
| \phi \rangle = \frac{1}{\sqrt{2\pi}} \left ( |z=1\rangle + e^{i \phi} |z=-1\rangle \right ) ,
\end{equation}
so that, the exact phase distribution is 
 \begin{equation}
   \label{PPe}
 P_\Phi (\phi) =  \langle \phi | \rho | \phi \rangle = \frac{1}{2 \pi} \left (1 +  \cos \phi s_x +   \sin \phi s_y \right ) ,
 \end{equation}
as a clear quantum counterpart of the classical $I_\Phi (\phi)$.

\subsection{Joint observation and inversion}

Thanks to the Hilbert-space formulation of classical interference, now the joint  observation of $\Phi$ and $Z$ follows exactly the same steps and settings of the classical case. We can proceed directly to the statistics for the so constructed joint observation of interference and polarization as
 \begin{equation}
 \tilde{P} (p, \phi)  =  \langle p | \langle \phi |  \rho |\phi \rangle | p \rangle ,
 \end{equation}
 where $| p \rangle$ are defined as in Eq. (\ref{|p>}), and $| \phi \rangle $ are in Eq. (\ref{ps2}),
 leading to 
\begin{eqnarray}
 & \tilde{P} ( p ,\phi )  = \frac{1}{4 \pi} \left [ 1 + \sin \left ( 2 \vartheta \right ) \left ( \cos \phi s_x + \sin \phi s_y \right ) \right . & \nonumber \\ 
 & \left . - z \cos \left ( 2 \vartheta \right )s_z \right ] .
 \end{eqnarray}
 
The observed marginal for $Z$ is 
\begin{equation}
\label{PZa}
 \tilde{P}_Z (z)  = \frac{1}{2} \left [ 1 - z \cos \left ( 2 \vartheta \right ) s_z \right ],
 \end{equation}
while the observed marginal for the phase is 
 \begin{equation}
   \label{PPa}
 \tilde{P}_\Phi (\phi) =  \frac{1}{2 \pi} \left [ 1 + \sin \left ( 2 \vartheta \right ) \left ( \cos \phi s_x + \sin \phi s_y \right ) \right ].
 \end{equation}
Then it is possible to obtain the exact  statistics $P_Z (z) $, $P_\Phi (\phi) $ in Eqs. (\ref{PZe}) and (\ref{PPe}) from the operational ones $\tilde{P}_Z (z) $, $\tilde{P}_\Phi (\phi) $ in Eqs.  (\ref{PZa}) and (\ref{PPa}) as 
\begin{equation}
\label{invZ}
P_Z (z) = \sum_{p} M_Z (z, p ) \tilde{P}_Z (p ) , 
\end{equation}
with the same $  M_Z (z, p )$ in Eq. (\ref{Mz}), while the phase distribution can be inverted as
\begin{equation}
\label{invP}
P_\Phi (\phi) = \int d \phi^\prime M_\Phi (\phi, \phi^\prime ) \tilde{P}_\Phi (\phi^\prime ) , 
\end{equation}
with the same $M_\Phi$ in Eq. (\ref{Mf}). Finally we extend the inversion (\ref{invZ}) and (\ref{invP})  from the marginals to the complete joint distribution 
to obtain a joint distribution $P  (z,\phi)$ as:
\begin{equation}
\label{tt0}
P  (z, \phi)  = \frac{1}{4 \pi } \left [ 1+ \cos \phi s_x + \sin \phi s_y + z s_z \right ] .
\end{equation}

\subsection{Pathology}

Let us examine whether $P(z, \phi)$ can take negative values. The minimun in Eq. (\ref{tt0})  is 
\begin{equation}
P_\mathrm{min}  = \frac{1}{4 \pi} \left ( 1 - |s_z | - \sqrt{ s_x^2 + s_y^2 } \right ) .
\end{equation}
Therefore,  $P_\mathrm{min}  < 0$  provided that  
\begin{equation}
\label{cond}
| \mu |^2 > \frac{1-|s_z|}{1 + |s_z|} =\min \left \{ P_Z (1)/P_Z (-1), P_Z (-1)/P_Z (1) \right \} ,
\end{equation}
where $| \mu |$ has exactly the same meaning as in the classical case
\begin{equation}
| \mu |^2 = \frac{|\rho_{1,-1}|^2}{\rho_{1,1} \rho _{-1,-1}} = \frac{s_x^2+s_y^2}{1-s_z^2} .
\end{equation}
So condition (\ref{cond}) for pathological behaviour is exactly the same than in the classical domain (\ref{PathologyClass}). This is clearly illustrated in Fig. 2. 
\begin{figure}[h]
    \centering
    \includegraphics[width=8cm]{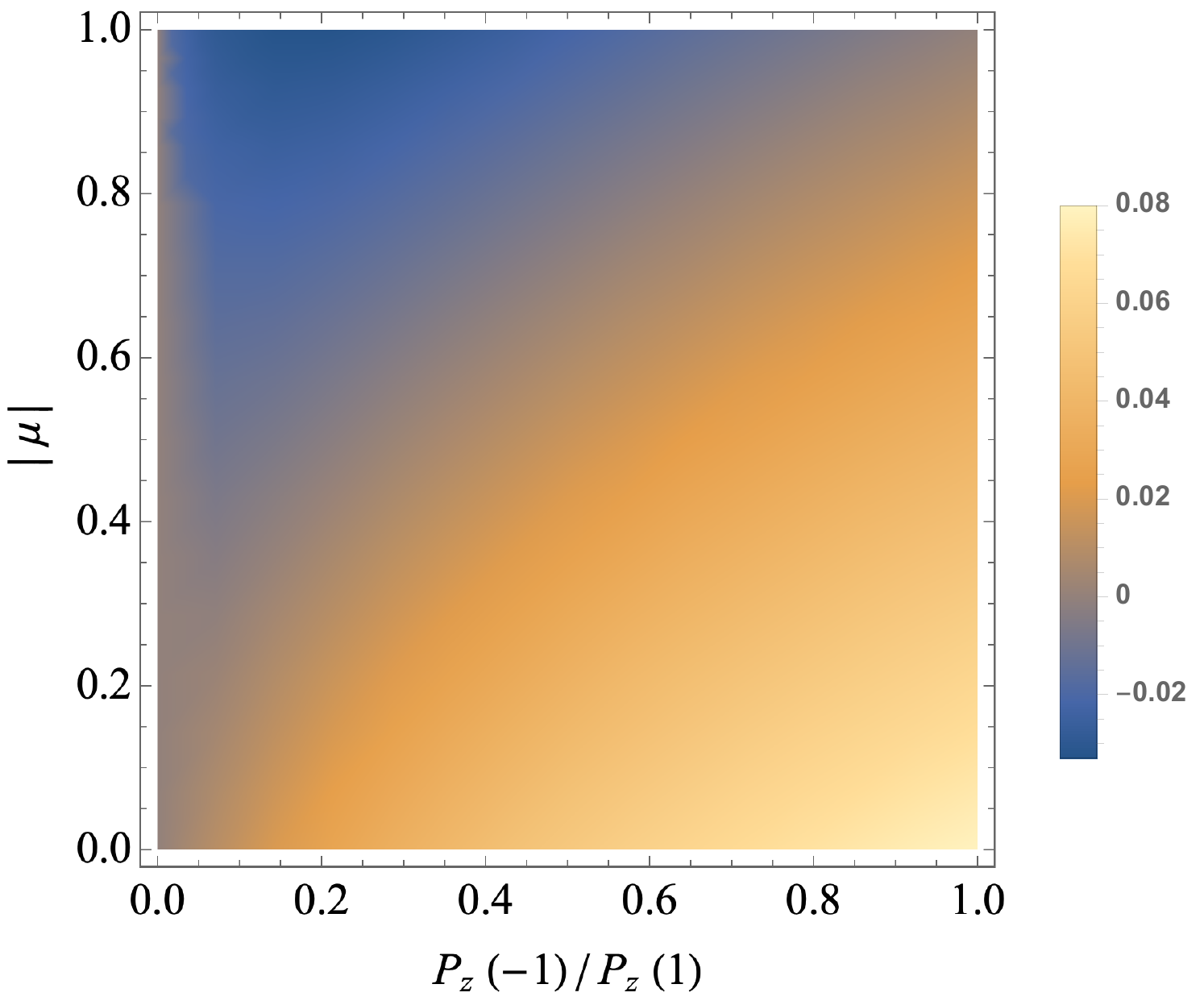}
    \caption{Retrieved joint probability distribution showing negative values under condition (\ref{cond}),  assuming $P_Z (-1) < P_Z(1)$ without loss of generality.}
    \label{figPpat}
\end{figure}{}

\section{Conclusions}

We have shown that there is a complete parallelism between quantum and classical complementarity. We have found that the classical joint intensity distribution exhibits negative values, and this negativity appears under the same conditions as it does in the probability distribution in the quantum scenario. 

A distinctive feature of the analysis proposed in this work is that we can use exactly the same tools in both domains. That highlights the structure of complementarity shared by the classical and quantum optics. So the approach presented in this work may provide a suitable arena for a deeper understanding of the quantum to classical transition.

 \bigskip

\section*{ACKNOWLEDGMENTS}

We thank Profs. X.-F. Qian and A. Goldberg for valuable comments. 
L. A. and A. L. acknowledge financial support from Spanish Ministerio de Econom\'ia y Competitividad Project No. FIS2016-75199-P, and from the Comunidad Aut\'onoma de Madrid research  consortium QUITEMAD+ Grant No. S2013/ICE-2801.
L. A. acknowledges financial support from European Social Fund and the Spanish Ministerio de Ciencia Innovaci\'{o}n y Universidades, Contract Grant No. BES-2017-081942. 

\bigskip

\appendix

\section{Inversion procedure. }

We consider two complementary observables $A$ and $B$, taking values $a$ and $b$. We propose the joint observation of two magnitudes $\tilde{A}$ and $\tilde{B}$, that can be considered as blurry counterparts of $A$ and $B$, respectively. Thus, after observation and measurement we get a well behaved operational joint distribution $\tilde{W} (a,b)$ with marginals 
\begin{equation}
\label{om}
\tilde{W}_A (a) = \sum_b \tilde{W} (a,b) , \qquad \tilde{W}_B (b) = \sum_a \tilde{W} (a,b) .
\end{equation}
We assume that these marginals provide complete information about the two observables $A$ and $B$. This implies that  there are functions $M_A (a, a^\prime )$ and $M_B (b, b^\prime )$  such that the exact distributions $W_A  (a)$ and $W_B  (b)$ can be retrieved from the observed marginals (\ref{om}):
\begin{eqnarray}
\label{inv}
& W_A (a) = \sum_{a^\prime} M_A (a, a^\prime )\; \tilde{W}_A  (a^\prime) , & \nonumber \\
 & & \\
 & W_B (b) = \sum_{b^\prime} M_B (b, b^\prime )\; \tilde{W}_B  (b^\prime) . \nonumber
\end{eqnarray}
The functions $M_A (a, a^\prime )$ and $M_B (b, b^\prime )$ are completely known since we assume that we know all the details about the measurement being performed.

The key idea is to extend this inversion (\ref{inv}) from the marginals to the complete joint distribution to obtain a joint distribution 
$W(a,b)$ for $A$ and $B$ as \cite{M1,M2,M3,M4,L1,L2,L3,L4}:
\begin{equation}
W (a,b)  = \sum_{a^\prime, b^\prime}\, M_A (a, a^\prime)\, M_B (b, b^\prime )  \,\tilde{W} (a^\prime, b^\prime ) .
\end{equation}

This distribution $W (a,b) $ is the one we expect to be pathological by taking negative values as a consequence of $A$ and $B$ being complementary. A key point is that this inversion protocol is exactly the same in quantum and classical optics. The only difference is the physical meaning of $W(a,b)$.

\end{document}